\documentclass[conference]{IEEEtran}

\usepackage{graphicx}
\usepackage{algorithm}
\usepackage{algpseudocode}

\IEEEoverridecommandlockouts
\IEEEpubid{%
   \makebox[\columnwidth]{ISBN 978-3-901882-83-8~\copyright~2016 IFIP \hfill}%
   \hspace{\columnsep}%
   \makebox[\columnwidth]{ }%
}

\begin{document}

\title{Multi-Flow Congestion Control with Network Assistance}

\author{
\IEEEauthorblockN{Yannis Thomas, George Xylomenos, Christos Tsilopoulos and George C. Polyzos}
\IEEEauthorblockA{Mobile Multimedia Laboratory \& Department of Informatics\\
School of Information Sciences and Technology\\
Athens University of Economics and Business\\
Patision 76, Athens 10434, Greece\\
E-mail: \{thomasi, xgeorge, tsilochr, polyzos\}@aueb.gr}
}

\maketitle

\begin{abstract}
A well-known technique for enhancing the performance and stability of content distribution is the use of multiple dissemination flows. Multipath TCP (MPTCP), the most popular multiflow protocol on the Internet, allows receivers to exploit multiple paths towards a single sender. Nevertheless, MPTCP cannot fully exploit the potential gains of multipath connectivity, as it must fairly share resources with (single-flow) TCP, without a clear understanding of whether the available paths do share any bottleneck links. In this paper, we introduce a hybrid congestion control algorithm for multisource and multipath transport that enables higher bandwidth utilization compared to MPTCP, while remaining friendly to TCP-like flows. Our solution employs (i) an in-network module that offers essential topological information and (ii) Normalized Multiflow Congestion Control (NMCC), a novel end-to-end congestion control algorithm. While NMCC is architecture-independent and the in-network module can be adapted for Multi-Protocol Label Switching (MPLS) or Software Defined Networks (SDNs), our prototype was implemented on the Publish-Subscribe Internetworking (PSI) architecture, which offers centralized path formation and source routing. Using an actual protocol implementation deployed on our test-bed, we provide experimental results which validate the effectiveness of our design in terms of performance, adaptation to shifting network conditions and friendliness to other flows.
\end{abstract}


%
\IEEEpeerreviewmaketitle

\section{Introduction}
\label{introduction}

Experience with content distribution indicates that multisource and multipath~\cite{1}, i.e. the use of multiple sources and multiple paths to each source, respectively, can benefit both network operators and end users. First, the exploitation of multiple paths allows achieving higher throughput via bandwidth aggregation. Second, the use of multiple sources offers resilience to both link and source failures via path or source switching. As a result, multisource and multipath, collectively referred to as \emph{multiflow} in this paper, provide load balancing and higher resource utilization, by spreading flows across more links and sources. 

Multiflow transport is the focus of considerable research activity, due to the increase of multihomed devices, such as smartphones with WiFi, Bluetooth and Cellular connectivity. A significant body of research has focused on the side-effects of multipath, such as lack of TCP friendliness~\cite{2,11,13,19,20,31}. This issue arises from the \emph{uncoupled} congestion control scheme originally proposed for \emph{Multipath TCP}~(MPTCP), where an independent congestion window is used for each subflow. This causes the multiflow transfer of $N$ flows to grasp up to $N$ times more bandwidth than a single-path flow over the same bottleneck, thus causing the latter to starve. The current MPTCP congestion control algorithm achieves TCP-friendliness by limiting all subflows, so as to fairly share bandwidth with single-path flows. Nevertheless, blindly restricting multipath flows can lead to degraded resource utilization when friendliness is not an issue, for example, when a multipath flow exploits physically disjoint paths. 

Efficient utilization of network resources is also the driving force of the \emph{Publish Subscribe Internet}~(PSI) architecture, an instantiation of the \emph{Information-Centric Networking}~(ICN) paradigm~\cite{3}. Following ICN principles, PSI bases communication on self-identified information items, rather than end-hosts. PSI also supports centralized path selection via a special network entity, the \emph{Topology Manager}, and source routing via LIPSIN forwarding~\cite{5}. We have exploited these features in previous studies~\cite{16,17}, where we presented the \emph{Multisource and Multipath Transfer Protocol}~(mmTP), a multiflow transport protocol for PSI. The use of multiple paths to multiple sources was shown to greatly enhance the performance and resilience of mmTP over the, inherently unpredictable, PlanetLab testbed~\cite{17}.

In this paper we focus on multiflow congestion control, proposing to exploit any available topological knowledge of the network to better balance performance and friendliness. Specifically, we present and evaluate a novel congestion control scheme for multiflow transport that consists of two independent modules: (i) \emph{Normalized Multiflow Congestion Control}~(NMCC), an end-to-end multiflow-aware algorithm, and (ii) an in-network mechanism to assist NMCC. NMCC is a simple, yet effective algorithm that manages bandwidth aggregation under the friendliness constraint, even in the face of heterogeneous paths and sudden changes in the congestion level. On the other hand, the in-network mechanism provides information about shared bottlenecks, thus allowing NMCC to adapt its behavior accordingly. Furthermore, we explain how our scheme can be adapted to IP networks operating over technologies utilizing centralized path computation components, including \emph{Multi-Protocol Label Switching}~(MPLS) and \emph{Software Defined Networks}~(SDNs).

The remainder of this paper is organized as follows. In Section~\ref{background} we summarize existing work on multiflow transport in IP and ICN networks. In Section~\ref{PSI} we briefly describe PSI and its features that allow us to realize selective friendliness. In Section~\ref{friendliness} we introduce our hybrid congestion control scheme, which consists of NMCC and the in-network assistance mechanism. In Section~\ref{implementation} we experimentally evaluate our design, using a prototype implementation. In Section~\ref{network_assisted} we explain how the required in-network mechanisms can be provided by MPLS and SDNs. We provide our conclusions in Section~\ref{conclusions}.

\section{Background Work}
\label{background}

Multipath congestion control is an active research topic for both traditional IP networks and ICN clean-slate architectures. The common goal is maximizing resource utilization, in terms of exploiting the bandwidth available in multiple paths, while not harming competitive single-flow transfers, a constraint also known as \emph{TCP-friendliness}.

\subsection{TCP-friendliness}

When a multiflow connection with $N$ independent subflows competes against a single-flow connection for the same bottleneck link, the multiflow connection can be up to $N$ times as aggressive as the single-flow one. While we usually say that the multiflow connection is not \emph{TCP-friendly}, we will use the term \emph{friendly} to imply \emph{single-flow friendly}, defined as follows:

\emph{When a multiflow connection competes with a single-flow connection for the same network resource, the former should not acquire a larger share of that resource than the latter.}

The price of friendliness is performance degradation: often, the bandwidth of the multiple subflows is not fully exploited, to prevent the starvation of single-flow connections. However, when the paths taken by each subflow are disjoint, meaning that we do not have multiple subflows sharing the same bottleneck link, this needlessly penalizes the multiflow connection. 

\subsection{Multiflow Congestion Control in IP}


The \emph{coupled\footnote{We use the term \emph{coupled} to refer to the final algorithm presented in~\cite{2}.}} congestion control algorithm of Multipath TCP (MPTCP) jointly tackles performance and friendliness~\cite{2}. MPTCP represents an evolution of TCP-Reno and EWTCP \cite{19}, adopting the slow-start and congestion avoidance phases per subflow, while also addressing multipath-specific problems, such as fair bottleneck sharing, \emph{Round Trip Time}~(RTT) mismatch and shifting network load. MPTCP manages its subflows under two constraints: (i) a multipath flow should achieve at least as much throughput as it would get with single-path TCP on the best of its paths and (ii) a multipath flow should grasp no more capacity on any path or collection of paths than a single-path TCP flow using the best of those paths. The second constraint, which assures MPTCP's friendliness towards unicast connections, compromises performance when friendliness in not an actual issue, for example, when the available paths do not share a bottleneck link. 

Even though this decision may be far from optimal, it is imposed by the IP routing architecture. Due to the distributed, hop-by-hop routing of IP networks, a transport protocol cannot reliably detect whether the dissemination paths used are overlapping. As a result, its congestion control module cannot detect whether friendliness is an issue or not. There are some application-layer solutions for the end-to-end detection of shared bottlenecks~\cite{32,33}, but their efficiency is debatable. In \cite{32} the authors detect shared bottlenecks based on the temporal correlation of fast-retransmit packets, while in \cite{33} the authors evaluate both loss-based and delay-based correlation techniques, arguing that the loss-based technique is unreliable, while the delay-based methods require considerably more time for accurate results; also, the convergence time of the loss-based method is roughly 15~ms, which is unrealistically high for a general purpose multiflow protocol. 

\subsection{Multiflow Congestion Control in ICN}

In ICN networks, the location-based networking of IP is replaced with information-based routing and forwarding. These features can support more efficient transport patterns, such as multipath, multisource and multicast, since they pin transport paths on the physical topology. Thereupon, it is often proposed that ICN routers
should assist topology-aware congestion control so as to better handle friendliness issues.

Along these lines, in~\cite{26} and~\cite{27} the authors discuss the design of transport protocols that  pull data from multiple sources via multiple paths over the \emph{Content Centric Networking}~(CCN) architecture~\cite{3}, exploiting congestion detection and control in the forwarding nodes. In~\cite{26} flow control and part of congestion control is managed by the receiver, but in-network congestion control is also present in the form of dynamic request forwarding: intermediate routers choose on-the-fly the most appropriate interface to forward each packet, shifting flows to less congested parts of the network. In contrast, in~\cite{27}, traffic control is exclusively assigned to in-network nodes, which separate content (cache) from forwarding (queue) storage: each router maintains a per-flow queue with the \emph{Deficit Round Robin}~(DRR) scheduling policy to determine which packets must be dropped and/or connections must be rejected, based on link utilization and fairness constraints. The receiver uses a simple control loop, responding to explicit congestion signals from routers.

The stateful CCN-based approaches have some important disadvantages. First, CCN nodes face significant overheads: the estimation of link utilization for congestion detection in~\cite{26,27} and the additional per packet state for fair queuing in~\cite{27} can impact their performance, making the achievement of wire speed forwarding doubtful. Second, distributed in-network congestion control has a delayed reaction to losses. While TCP rapidly detects lost packets via either out-of-sequence packets or time-outs, in~\cite{27} authors use explicit notifications to the receiver when a queue drops a packet; \cite{26} introduces a novel time-out estimation function, which is not investigated with regard to its effects on the other CCN timers. 

A different approach for enriching congestion control with topological information, involves an in-network notification system that can report the existence of shared bottlenecks. This notification system, which must be aware of both network structure and dissemination routes, should explicitly indicate path disjointness to the end-hosts, allowing them to apply friendliness mechanisms more selectively. This design offers accurate information without convergence delay and without stressing the core routers, which are the weaknesses of the IP and CCN solutions, respectively. PSI follows this approach, since routing takes place at a conceptually centralized in-network entity, the \emph{Topology Manager}. We briefly discuss the PSI architecture in the following section.

\section{Multiflow transport in the PSI architecture }\label{PSI}

\subsection{The PSI architecture}

In the PSI architecture, content objects are treated as publications, content sources as publishers and content consumers as subscribers. User programs exploit a publish/subscribe API for advertising and requesting information. A fundamental design tenet in PSI is the clear separation of its core functions~\cite{4}: (i) the Rendezvous function tracks available publications and resolves subscriptions to publishers, (ii) the Topology Management and Path Formation function monitors the network topology and forms forwarding paths and (iii) the Forwarding function handles packet forwarding~\cite{commag}. 

Network nodes in a PSI network are classified into \emph{Rendezvous Nodes}~(RNs), \emph{Topology Managers}~(TMs) and \emph{Forwarding Nodes}~(FNs). The RNs receive and store the pub/sub requests and match publications with subscriptions of the same content. 
When matching takes place, the RN asks a TM to find the appropriate dissemination routes. The TM, which is aware of topology, network conditions and content characteristics, discovers the ``best'' path(s) and encodes them into LIPSIN identifiers~\cite{commag}. LIPSIN forwarding, which is realized by the FNs, offers line-speed stateless source routing. Finally, the LIPSIN identifiers are delivered to the end-host applications that exploit them for direct communication, thus delegating congestion control to the network edges.

The centralized nature of the TMs raises concerns about PSI's feasibility, since they must compute paths for all network connections. However, recent work showed that a centralized intra-domain TM service is feasible: for a typical national-scale network provider in the UK, it was demonstrated that a reasonable number of TM instances with precomputed paths can efficiently cope with the resulting network load~\cite{martin}.

\subsection{Multipath and multisource in PSI}\label{mmTP}

We have presented a multiflow transport protocol for PSI in previous studies~\cite{16,17}, the \emph{Multisource and Multipath Transfer Protocol}~(mmTP).
mmTP is a reliable protocol that supports multisource and multipath data transfers by exploiting PSI's source routing and centralized path selection. mmTP relies on a TM function that can discover multiple paths between a receiver and multiple senders. These paths are encoded in LIPSIN identifiers that are later sent to the end-hosts. Given that LIPSIN identifiers encode dissemination routes without unveiling the actual dissemination paths, or even the destination nodes, the end-hosts acquire a set of  distinct “options” for requesting data, which may involve different publishers and/or different paths. Hence, mmTP provides a generic interface, transparently supporting any combination of multisource and/or multipath services. 

The design of mmTP allows congestion control in two levels: (i) path selection by the TMs and (ii) path utilization by the end-hosts. Specifically, the TMs, which are aware of network conditions, select appropriate routes for load balancing and bandwidth aggregation. We have previously shown the gains of centralized path formation in \cite{34}, where we used QoS routing schemes to satisfy certain throughput and error rate constraints in PSI. Based on these routes, the end-hosts evaluate in real-time the performance of each path and adjust the amount of data to be delivered through it. The congestion control mechanism used at the end-hosts, which is derived from TCP, pushes complexity at the network edges, thus enhancing network stability and keeping forwarding stateless.

\begin{figure*}[htbp]
\centering
\includegraphics[scale=0.5]{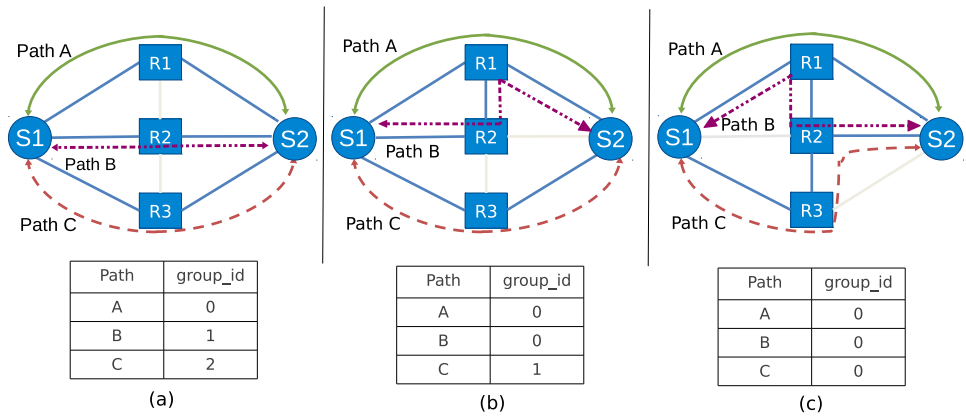}
\caption{Three different cases of path composition and their corresponding $group\_id$ codes: (a) Disjoint paths, (b) Paths A and B share one link, (c) Paths A and B share one link, paths B and C share another link.}
\label{path_form}
\end{figure*}

\section{Hybrid multi-flow congestion control}
\label{friendliness}

In this section we present a hybrid multiflow congestion control algorithm that enhances resource utilization without violating the friendliness requirement. 
Our novel congestion control scheme consists of two independent modules: (i) NMCC, an end-to-end multiflow-aware algorithm, and (ii) an in-network mechanism to assist congestion control. NMCC is simple, yet it outperforms the coupled congestion control of MPTCP in terms of friendliness in short transfers and performance in heterogeneous networks. The in-network mechanism exploits knowledge of shared bottlenecks to enhance the performance adaptation of NMCC.

\subsection{Path Formation}
\label{Path_Formation}

The best case scenario for multiflow communication arises when all communication paths are physically disjoint, that is, they do not share \emph{any} links or routers. In this case, each multiflow connection can use the same congestion control algorithm as single-flow connections. In contrast, when some subflows use paths which are not disjoint, their aggressiveness needs to be limited in order for them to remain friendly.

Path selection in PSI is performed by the TMs, whose operation extends beyond the scope of this paper. Our only requirement is that when the TMs return a set of paths encoded as LIPSIN identifiers, a $group\_id$ code should be added to each identifier so as to indicate non-disjoint paths. Specifically, all paths that share at least one link with each other (not necessarily \emph{the same} link) are marked with the same $group\_id$. In general, for any given underlying routing mechanism, the in-network assistance mechanism must be able to signal to NMCC how the available paths are grouped by $group\_id$.

For example, Figure~\ref{path_form} shows three examples of path composition along with the corresponding $group\_id$ codes. In Figure~\ref{path_form}(a) the three paths are disjoint, thus each path is marked with a distinct $group\_id$, whereas in Figure~\ref{path_form}(b) paths A and B share a link, thus they have the same $group\_id$. In \ref{path_form}(c) Paths A and B share a link and paths B and C share a different link; they still get the same $group\_id$, to ensure that each path belongs to a single group. This simplifies operation, at the cost of losing some efficiency, since a congested link may only affect some of the paths in a group. 

\subsection{Window Management}

When the available paths have different $group\_ids$ (i.e., they do not share any links), then window management does not consider friendliness: our algorithm creates a distinct TCP-like subflow for each path with an individual congestion window variable ($cwnd$), RTT-based loss detection mechanism, retransmission mechanism and \emph{slow start} and \emph{congestion avoidance} algorithms. Therefore, window management is similar to MPTCP's uncoupled congestion control scheme.

%

In contrast, when multiple paths have the have the same $group\_id$ (i.e., they share some links) our NMCC algorithm is used to maintain friendliness. 
NMCC differs from the coupled congestion control algorithm of MPTCP in two respects. 
First, coupled MPTCP only tries to limit its aggressiveness during congestion avoidance, while NMCC also considers slow start. 
Second, NMCC is simpler to operate than coupled MPTCP. The coupled MPTCP algorithm is a variant of TCP Reno, where aggressiveness is controlled by reducing the growth rate of the congestion window per RTT. This introduces complications when paths have different RTTS and shifting network loads, which are due to the use of a single-flow solution to a multiflow problem. On the other hand, NMCC exploits a well-known TCP-fairness issue, the fact that connections with higher RTTs are less aggressive~\cite{15}, to ensure friendliness. Instead of reducing the growth of the congestion window per RTT, NMCC controls the congestion window by inflating the RTTs; this reduces complexity, simplifies friendliness during slow start and avoids multiflow-related issues due to RTT mismatch and sudden load and congestion shifts.


\subsubsection{Congestion Avoidance}

NMCC uses an inflated $RTT'_i \geq RTT_i$ for each subflow $i$ to control window growth; the inflated $RTT'_i$ makes the congestion window grow slower compared to a single-flow connection. We introduce a \emph{friendliness factor} $m \geq 1$ so that $RTT'_i=m*RTT_i$, trying to approximate the two goals of fair bottleneck sharing: (i) the growth rate of all subflows sharing a link should be no more than that of a single-flow connection and (ii) the overall growth rate should not be less than that of the most aggressive single-flow connection. Since the most aggressive single-flow connection has the minimum $RTT_i=RTT_{min}$ and during congestion avoidance the growth rate of a single-flow connection is one packet per $RTT$, the rate increase intervals during congestion avoidance must satisfy the following equation:
\[ \frac{1}{RTT_{min}} = \sum_{i=1}^N \frac{1}{RTT'_i} = \sum_{i=1}^N \frac{1}{m*RTT_i}\]
where $N$ is the number of jointly controlled subflows. We can therefore estimate $m$ using the following equation:
\[ m = RTT_{min} * \sum_{i=1}^N \frac{1}{RTT_i} \]

To understand the friendliness factor $m$, consider a simple example. When the TM offers two paths marked with the same \textit{group\_id}, we initially set $m=2$, the number of jointly controlled paths. Upon receipt of the first packet over each path, the $RTT_i$'s are updated and $m$ is re-calculated. If $RTT_1 = 50$~ms and $RTT_2 = 100$~ms, then $m = 1.5$, so $RTT'_1 = 75$~ms and $RTT'_{2} = 150$~ms, therefore NMCC will increase its overall congestion window by three \emph{maximum segment sizes}~(MSS) during a period of $150$~ms: two MSS from the first subflow and one MSS from the second. This is equal to the increase of the fastest single-flow connection: one MSS per $50$~ms.

By applying $m$ to all subflows, we adapt the growth rate across all paths. This means that, although we favor the subflow which operates over the fastest path, we do not neglect the slower paths. Therefore, NMCC does not require probing to detect load changes on an unused path, whereas the coupled MPTCP algorithm introduces a special parameter for controlling the amount of probing. NMCC can therefore perform efficiently in heterogeneous environments, adapting fast to path failures and congestion bursts. For instance, consider an integrated terrestrial-satellite network where the terrestrial link has 10~ms delay and the satellite one 250~ms. In this case $m=1.004$, hence window growth is not constrained and NMCC effectively grasps the available capacity.

\subsubsection{Slow Start}
\label{Slow_Start}

Most work on multiflow transport only deals with congestion avoidance, disregarding slow start. Nevertheless, during the evaluation of NMCC we noticed that friendliness was compromised when (i) the content was relatively small and (ii) the path was very congested. An analysis of the evolution of the congestion windows showed that NMCC gained bandwidth almost $N$-times faster than a single-flow connection during slow start, with $N$ subflows. Since short and very congested connections spend a substantial fraction of their lifetimes in slow start, meeting the friendliness goals in congestion avoidance was not enough to amortize NMCC's aggressive behavior during slow start.

One way to reduce aggressiveness during slow start is to reduce $ssthresh$ so as to move faster to congestion avoidance. Unfortunately, this has two disadvantages. First, when a connection starts, the available bandwidth of the communication path is unknown, thus $ssthresh$ should be set high enough to probe it. Second, reducing $ssthresh$ only limits the amount of bandwidth that the protocol will re-acquire, not its rate of acquisition. For this reason, we reused the friendliness factor $m$ to also control slow start.

During slow start, a subflow $i$ doubles its congestion window during a period of $RTT_i$; its growth rate is $\frac{cwnd_i}{RTT_i}$, while during congestion avoidance it drops to $\frac{1}{RTT_i}$. We introduce $\Omega_i$ and $\Omega'_i$, the regular and the friendly growth rate of subflow $i$, respectively, where $\Omega_i = m * \Omega'_i$. Again, we want to match the growth rate of the most aggressive flow, $\Omega_{max}$, therefore we have the following equation:
\[ \Omega_{max} = \sum_{i=1}^N \Omega'_i = \sum_{i=1}^N \frac{\Omega_i }{m}\]
for $N$ jointly controlled subflows. We can then calculate $m$ based on the regular growth rates of all subflows as follows:
\[ m = \frac{\sum_{i=1}^N \Omega_i }{\Omega_{max}} \]
Consequently, each flow's growth rate $\Omega'_{i}$ becomes $\frac{cwnd^{tcp}_i}{m*RTT_i}$ during slow start and $\frac{1}{m*RTT_i}$ during congestion avoidance, where $cwnd^{tcp}_i$ is the equilibrium window of TCP for path $i$. 
As increases in slow start are multiplicative, any change in window growth affects the subsequent increases: smaller windows grow slower. Therefore, during slow-start we use $cwnd^{tcp}$ in order to assure that the cumulative growth of NMCC is equal to single-flow TCP.
Algorithm~\ref{window_incr_algo} provides the combined slow start and congestion avoidance algorithm. Note that the algorithm translates the ``inflated RTTs'' of NMCC into MPTCP-like ``decreased window growths'' to avoid any side-effects of prolonged timeouts, such as delayed loss detection.

\begin{algorithm}
\caption{Window adjustment and estimation of $m$.}\label{window_incr_algo}
 \begin{algorithmic}[1]
  \Procedure{increase\_window}{}
 		\If {($cwnd < ssthresh$)}
 		  \State $cwnd \gets cwnd + cwnd^{tcp} * MSS / (cwnd * m)$ 
 		\Else
 		  \State $cwnd \gets cwnd + MSS / (cwnd * m)$
 		\EndIf
  \EndProcedure
 \end{algorithmic}  
 \begin{algorithmic}[1]
 \Procedure{estimate\_m}{}
		\State $max\_rate \gets 0$
		\State $total\_rate \gets 0$
		\For{$(i \in subflows)$}
	 		\If {($subflow\_state_i == \mathrm{CONG\_AVOID}$)}
 			  \State $rate \gets MSS/RTT_i$
 			\Else
 		  	\State $rate \gets cwnd_i/RTT_i$
 			\EndIf
 			\State $total\_rate \gets total\_rate+rate$
 			\If {($rate > max\_rate$)}
 		 	 \State $max\_rate \gets rate$
 		 	
 		 	\EndIf
 		\EndFor 
 		\State $m \gets total\_rate / \max\_rate$
  \EndProcedure 
  \end{algorithmic}
\end{algorithm}

\section{Performance Evaluation}
\label{implementation}

In this section we focus on the extent to which our hybrid congestion control can meet the friendliness requirement of multiflow transfers in different network scenarios. We have implemented our scheme as part of the mmTP protocol that runs over Blackadder, the PSI prototype implementation~\cite{6}. Our implementation includes the mmTP sender and receiver applications with NMCC enabled, as well as a TM that computes the $k$-shortest paths from every publisher to a subscriber, using the algorithm by Yen~\cite{12} with hop count as the metric.\footnote{Our implementation is available at \texttt{http://mm.aueb.gr/}.} 

We deployed Blackadder with mmTP in several LAN topologies, using 100~Mbit switches and workstations as network nodes. 
Our experiments examine (i) the effect of TM assistance when paths are disjoint, (ii) the effectiveness of NMCC with overlapping paths, (iii) NMCC's behavior in short transfers (iv) the friendliness of NMCC and coupled MPTCP and (v) NMCC's behavior in heterogeneous networks. 

In our testbed, the transmission latency among all nodes is 0.2-0.3~ms and the bandwidth of each link is 11.7 MB/s, as estimated using \texttt{iperf}.\footnote{Available at \texttt{http://iperf.sourceforge.net/}.} The duration of transfers during all experiments is 20 seconds, except when mentioned otherwise. In order to enhance the reliability of our conclusions, we repeated each experiment until the margin of error was less than 1\%, so as to achieve a confidence level of 95\%.

\begin{figure*}
\centering
\includegraphics[scale=0.45]{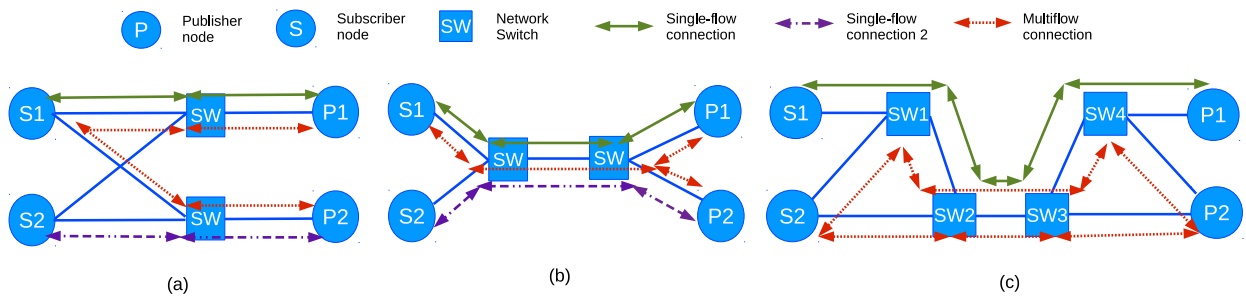}
\caption{Topology for performance evaluation with (a) disjoint paths and (b)-(c) shared paths.}
\label{lan_top}
\end{figure*} 

\subsection{Disjoint paths}
\label{TMs_assistance_in_cong_management}

We first deployed mmTP in the topology of Figure~\ref{lan_top}(a), where we investigated the performance gains of our hybrid congestion control scheme when paths are known to be disjoint. Figure~\ref{lan_top}(a) supports one multisource path from publishers P1 and P2 to subscriber S1 and two single-paths from publishers P1 and P2 to subscribers S1 and S2, respectively.
Thereupon, we ran some experiments with no contending traffic, so as to establish a performance baseline, leading to the average transfer rates shown in Table~\ref{res_disjoint_paths_isolated}. These experiments include deployment of multiflow mmTP connections with and without TM assistance, as well as single-flow mmTP connections. We notice that each path offers roughly 10.6 MB/s throughput and multiflow mmTP achieves 21.3 and 20.7 MB/s with and without TM assistance, respectively. These preliminary results validate that mmTP fully exploits available capacity and imply that TM assistance slightly enhances performance, even in the absence of competitive flows.

\begin{table}
\begin{center}
\begin{tabular}{ l r }
 \hline            
 {\bf Transmission mode} & {\bf Transfer rate (MB/s)} \\
 \hline  
 Multipath with TM assistance & 21.3 \\
 \hline
 Multipath with no TM assistance & 20.7 \\
 \hline 
 Single-flow from P1 to S1 & 10.6 \\
 \hline 
 Single-flow from P2 to S2 & 10.7 \\
 \hline 
 Single-flows on both paths & 21.1	\\ \hline
\end{tabular}
\end{center}
\caption{Average transfer rates with disjoint paths.}\label{res_disjoint_paths_isolated}
\end{table}

\begin{figure}[t]
\centering
\includegraphics[scale=0.5]{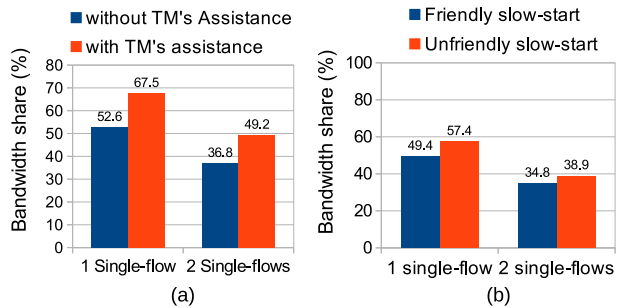}
\caption{Bandwidth share of mmTP (a) with and without TM assistance and (b) with and without friendly slow start in short transfers.}
\label{tm_assistance-short_transfers}
\end{figure}

We then deployed mmTP in multipath mode over the same topology (S1 requests data from both P1 and P2), with one or two single-flow connections competing over one or both disjoint paths (S1 to P1 and S2 to P2). In Figure~\ref{tm_assistance-short_transfers}(a) we show the average share of the total bandwidth that mmTP achieved in each case, depending on whether TM assistance was turned on or off. The results validate the performance gains and the friendliness of NMCC. Ideally, with one contending single-flow connection NMCC should use half of the bandwidth over one path and the entire bandwidth over the other, or 75\% of the total bandwidth, while with two contending single-flow connections NMCC should use half of the bandwidth over each path, or 50\% of the total bandwidth. With TM assistance, mmTP acquires 67.5\% and 49.5\% of the overall bandwidth, respectively. Not only is this higher than with no TM assistance, it is also closer to the ideal bandwidth share. The bandwidth shares of mmTP with no TM assistance, which are only 52.6\% and 36.8\%, respectively, correspond to an equal share of the bandwidth among all connections, disregarding the actual topology.

\subsection{Shared paths}

To investigate the case where paths share some links, mandating a less aggresive behavior to ensure friendliness, we used the topology shown in Fig.~\ref{lan_top}(b), where Publishers and Subscribers are connected by paths sharing a link. We deployed a multisource connection from subscriber S1 to publishers P1 and P2, in parallel with 1, 2, 4 and 9 single-flow connections from subscriber S1 to publisher P1 and from subscriber S2 to publisher P2; these connections are distributed uniformly between the two paths. 

\begin{figure*}[t]
\centering
\includegraphics[scale=0.45]{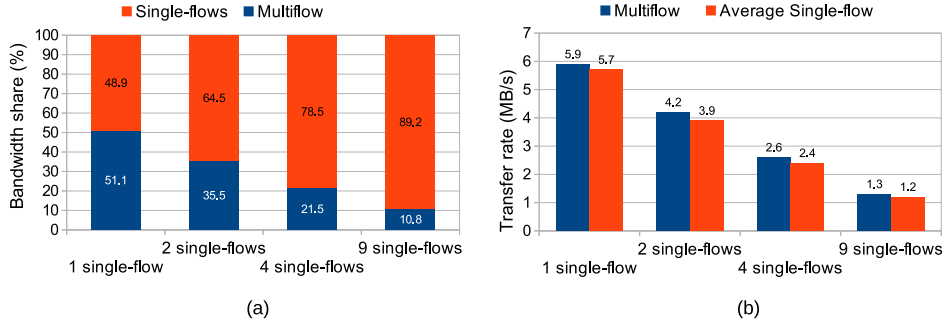}
\caption{(a) Bandwidth shares of NMCC and all single-flow connections. (b) Transfer rate of NMCC and the average single-flow connection.}
\label{shared_link}
\end{figure*}

Fig.~\ref{shared_link}(a) demonstrates the average bandwidth percentage acquired by NMCC and \emph{all} single-flow connections, while Fig.~\ref{shared_link}(b) displays the average transfer rate achieved by NMCC and the \emph{average} unicast connection. NMCC acquires 51.1\%, 35.5\%, 21.5\% and 10.8\% of the bottleneck link's bandwidth when competing with 1, 2, 4 and 9 single-flow connections, respectively, marginally over the optimal sharing ratios of 50\%, 33.3\%, 20\% and 10\%, respectively, thus satisfying the friendliness goal. The slight performance advantage of NMCC, also evident in the transfer rates, is a side effect of the friendliness constraint: since window growth is distributed across all subflows, NMCC approaches congestion limits gradually, resulting in slightly less retransmissions than the average single-flow connection (2.1\% on average).

We also examined NMCC's response to a sudden change in the congestion level, by repeating the previous experiment, but this time starting the multiflow connection either 7~sec after or 7~sec before the start of the single-flow connections. The results of these experiments are nearly identical to the previous ones, as NMCC acquires 52\%, 35.6\%, 21.1\% and 10.8\% of the bandwidth when competing with 1, 2, 4 and 9 single-flow connections, respectively. Consequently, NMCC manages to efficiently share bandwidth with newly established connections, as well as to obtain a fair share of bandwidth when launched in an already congested path. 

%

\subsection{Short Transfers}
\label{FriendlinessSlow-Start}

NMCC is friendly during slow-start, unlike MPTCP which is only concerned with congestion avoidance. This is particularly important for short transfers, where friendly congestion avoidance cannot compensate for an unfriendly slow start. To evaluate this aspect of NMCC, we reused the shared link topology of Fig.~\ref{lan_top}(b), deploying one multisource NMCC connection and either 1 or 2 contending single-flow connections. Each connection transfers a 10~MByte object, which would require less than 1.1~second to complete in the absence of contention. Fig.~\ref{tm_assistance-short_transfers}(b) presents the percentage of overall bandwidth acquired by NMCC when friendly slow start is turned on or off. 

With unfriendly slow start, NMCC grabs a disproportionate amount of bandwidth from the competing connections, compared to the ideal shares of 50\% and 33\%. In the first case, NMCC gets 57.4\% of the bandwidth; while in the second case it gets 38.9\%, or 14.8\% and 16.8\% more than the fair share, respectively. On the other hand, NMCC with friendly slow start gains 49.4\% and 34.8\% of the total bandwidth. Consequently, NMCC is friendly even with short transfers. 

For even shorter transfers, for example Web objects a few KBytes long, the unfairness is even more pronounced, as such connections can easily complete during slow start. The reason for presenting results from a 10~MByte transfer is to show that the initial over-aggressiveness during slow start cannot be compensated even with longer transfers.

\subsection{Friendliness of NMCC and MPTCP}

We then compared the friendliness of the hybrid approach of NMCC and the coupled congestion control of MPTCP~\cite{2}. MPTCP's design is similar to NMCC, in that congestion management takes place at the endpoints and time-out estimation is based on RTTs. These similarities simplified the implementation of the coupled congestion control algorithm of MPTCP in our mmTP implementation. 

For these experiments we used the topology of Fig.~\ref{lan_top}(c), where all paths share at least one link; MPTCP's inability to support TM assistance would make a comparison over disjoint paths unfair. We deployed a number of single-path flows from subscriber S1 to publisher P1, as well as multipath flows from subscriber S2 to publisher P2, using the paths indicated in Fig.~\ref{lan_top}(c). Multipath connections utilized either the coupled MPTCP or the NMCC algorithm. We denote each experiment as $X:Y:Z$, where $X$ shows the number of single-path flows, $Y$ shows the number of multipath flows using coupled MPTCP and $Z$ shows the number of those using NMCC. 

\begin{figure}[htbp]
\centering
\includegraphics[scale=0.35]{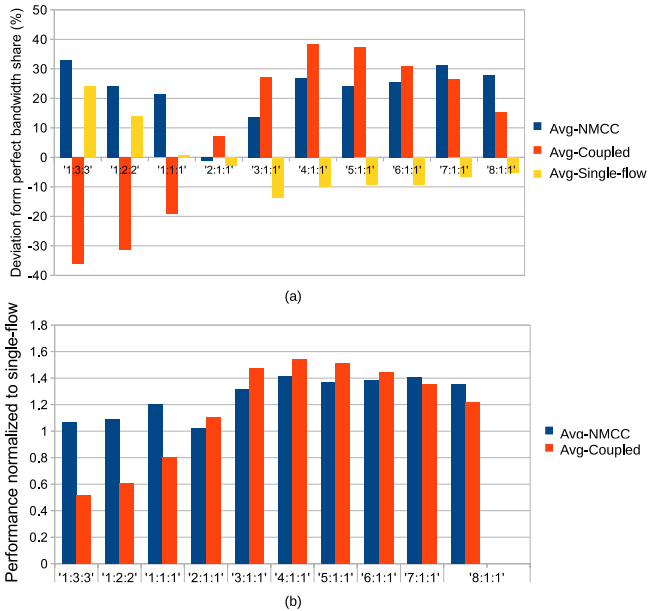}
\caption{(a) Deviation of obtained bandwidth from fair shares, (b) Difference of multiflow and single-flow deviation from fair shares normalized to the single-flow deviation from fair shares.}
\label{comparison_mtcp}
\end{figure}

The results of these experiments are summarized in Fig.~\ref{comparison_mtcp}. Fig.~\ref{comparison_mtcp}(a) displays the deviation of the obtained bandwidth of each connection from its fair share which, due to the shared link, is given by $\frac{Link\_Capacity}{\#Connections}$. Results below $0\%$ indicate overly friendly flows, while results over $0\%$ indicate overly aggressive ones. We can distinguish three groups in this figure. The first group reflects experiments `1:3:3', `1:2:2' and `1:1:1', where MPTCP is too friendly, resulting in poor performance. The second group reflects experiment `2:1:1', where all connections are close to their fair shares. The third group reflects experiments from `3:1:1' to `8:1:1', where multiflow connections are more aggressive, making single-flow ones lose some of their share.
 
Figure~\ref{comparison_mtcp}(b) presents the above results for multipath connections normalized to the bandwidth achieved by single-path flows, that is, we divide the bandwidth share obtained by MPTCP and NMCC by the bandwidth achieved by the average single-flow connection. Thereupon, the closer the score is to $1$ the friendlier a connection is to single-flow. Based on this figure we can argue that NMCC is more friendly to single-flow connections than MPTCP most of the time. Even though the performance superiority of NMCC is mostly evident when there are fewer single-path flows competing for capacity, we observe that NMCC gives more consistent results in general.

\subsection{Heterogeneous Networks}

\begin{figure}[htbp]
\centering
\includegraphics[scale=0.6]{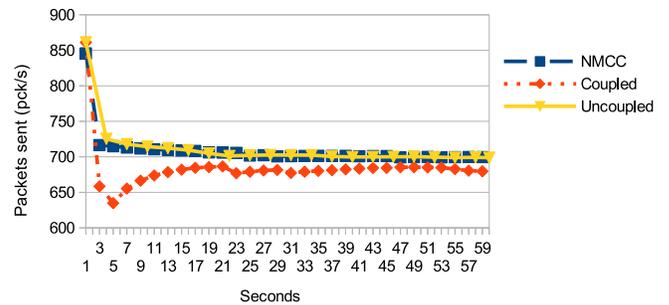}
\caption{Packets/sec sent over the WiFi link for a 60~sec period.}
\label{heterogeneous_res}
\end{figure}

Finally, we explored NMCC's performance in heterogeneous networks where paths exhibit diverse capacity, delays and error-rates.
In order to emulate these conditions, we replicated the RTT-mismatch scenario used in the evaluation of coupled MPTCP~\cite{2}. 
This scenario assumes a smartphone device that uses simultaneously two disjoint paths: (a) a WiFi link with 10~ms delay and 4\% error-rate and (b) a 3G link with 100~ms delay and 1\% error-rate. 
First, we used netem\footnote{http://www.linuxfoundation.org/collaborate/workgroups/networking/netem} to configure the delay and error-rate of the multisource paths in the topology of Figure~\ref{lan_top}(a) and then we deployed mmTP with no contending traffic, so as to study window growth without congestion.
We investigated the behavior of NMCC against both the coupled and uncoupled MPTCP congestion control algorithms. Figure~\ref{heterogeneous_res} presents the number of packets that are sent over the WiFi link within a period of 60 seconds; we neglect the 3G link, as it is identically saturated by all algorithms. The results validate the expected  performance superiority of NMCC. 
The significant RTT divergence leads NMCC to compute a low friendliness factor ($m \simeq 1.1$) which offers similar performance to the uncoupled MPTCP algorithm, thus grasping all available capacity from the start. In contrast, coupled MPTCP fails to adapt to this RTT mismatch, as it utilizes less than 93\% of the available capacity until 10~sec and roughly 96\% thereon.

\section{In-network assistance in IP networks}
\label{network_assisted}

Our hybrid congestion control mechanism for multiflow transfers, NMCC, relies on an in-network scheme that reports shared bottlenecks to the end-hosts. The PSI architecture is an appropriate terrain for this design, since it provides a TM function that (i) is aware of network topology and (ii) interacts with the end-hosts. The TM knows the physical structure of the network, so it can easily detect shared bottlenecks. In addition, when two pub/sub requests are matched, the TM sends the LIPSIN identifiers of the paths directly to the applications, therefore it directly pushes the topological information to the users. In order to extend our scheme to other types of networks, such as IP-based ones, we need equivalent in-network mechanisms to provide such information.

A technology that offers centralized path selection and source routing in IP networks is \emph{Multi-Protocol Label Switching}~(MPLS)~\cite{36}. MPLS is used in backbone networks, where it applies QoS-based traffic control by classifying flows and forwarding them via predefined routes. Short fixed-length labels are assigned to packets at the ingress to an MPLS cloud, and these labels are used to make forwarding decisions inside the MPLS domain. The path formation process is generic, allowing route computation by the underlying routing protocols or explicit definition by a network operator. Multipath deliveries are also encouraged, in the form of splitting single-flow connections into several subflows at the ingress router. 

Currently, MPLS is used for applying domain-scale traffic engineering rather than for enhancing the performance of individual connections, hence, connection splitting is done with static sharing weights for general load balancing. Consequently, congestion control takes place at the actual end-hosts (i.e the users), while the ingress MPLS router is confined to the flow control of the available paths. However, if we consider the ingress router as the congestion manager of the MPLS cloud, as it splits the flow, assigns labels to each of its subflows and becomes the end-host of a local MPLS service, then our network-assisted congestion control can be integrated to the MPLS network. Specifically, when the network operator discovers multiple paths for bulk flows and sends the corresponding labels to the ingress router, it also sends information on how flows are grouped depending on path sharing, as described in Sec.~\ref{Path_Formation}. The ingress router, which runs NMCC for each bulk flow, exploits this information and source routing to selectively engage the friendliness mechanism. 

\emph{Software-Defined Networking}~(SDN)~\cite{35} is a novel networking scheme that can be used to achieve similar goals to PSI, including centralized path selection. In SDN, programmable switches forward packets based on ``dynamic'' rules that bind flow identifiers, such as fields of the IP header, with outgoing network interfaces. These rules are defined by a centralized controller that is aware of the network topology and forms virtual circuits by explicitly sending rules to all on-path routers. Circuit creation can be reactive, where a router ask the controller's assistance when no rule can be applied to a received packet, or proactive, where the controller forms the route a priori, for example, to achieve load balancing. In both cases, SDN operation is transparent to the end-hosts that manage congestion control. 

As the SDN controller does not communicate with end hosts, which means that it cannot pass any topological information to them, we can apply the same ideas as for MPLS to introduce in-network assistance and NMCC to SDN clouds, by considering the ingress SDN router as the congestion manager of bulk flows. When the SDN controller creates forwarding paths by sending the appropriate rules to the SDN switches, it can send information on how flows are grouped depending on path sharing to the ingress SDN router, as well as instructions on how to tag each IP header so as to implicitly select the appropriate path. The ingress SDN router will then run NMCC for each bulk flow, as above. 

Adding in-network assistance to MPLS or SDN clouds may raise two concerns: (i) the computational costs of applying congestion control for numerous flows at the ingress router may degrade scalability, (ii) the limited application scope of backbone networks may prevent fully exploiting all connectivity options. For example, when multihomed devices connect to different access networks, these may not employ the same MPLS or SDN cloud, preventing the transparent use of NMCC within each separate cloud.

\section{Conclusions}
\label{conclusions}

We presented a hybrid congestion control algorithm for multiflow transport, consisting of NMCC and an in-network assistance mechanism. Our design offers friendliness to single path connections using TCP-like congestion control, while increasing the utilization of network resources. It achieves this by detecting shared physical bottlenecks and managing aggressiveness appropriately, without requiring complex network signaling or adding state to routers. We have implemented the congestion control algorithm in the PSI architecture prototype and evaluated its performance gains in several topological and traffic scenarios. Our results not only verify the effectiveness of our design, they also validate its performance superiority over MPTCP's coupled congestion control algorithm in short transfers and heterogeneous networks. Finally, we  discussed how in-network assistance can be provided in IP networks based on centralized routing, such as MPLS or SDN.

\section*{Acknowledgement}
The work presented in this paper was supported by the EU funded H2020 ICT project POINT, under contract 643990.


\end{document}